# A Directly Public Verifiable Signcryption Scheme based on Elliptic Curves [†]

Mohsen Toorani [‡]

Ali A. Beheshti

## Abstract

*A directly public verifiable signcryption scheme is introduced in this paper that provides the security attributes of message confidentiality, authentication, integrity, non-repudiation, unforgeability, and forward secrecy of message confidentiality. It provides the attribute of direct public verifiability so anyone can verify the signcryption without any need for any secret information from the corresponding participants. The proposed scheme is based on elliptic curve cryptography and is so suitable for environments with resource constraints.*

## 1. Introduction

The signcryption is a cryptographic technique that tries to fulfill the functionalities of digital signature and encryption in a single logical step, and decreases the computational costs and communication overheads in comparison with the traditional sign-then-encrypt schemes. Several signcryption schemes [1-7] are proposed over the years, each of them providing different level of security services and computational costs.

A signcryption scheme should simultaneously provide the security attributes of encryption and digital signature that mainly include: *Confidentiality, Unforgeability, Integrity*, and *Non-repudiation*. Some signcryption schemes provide further attributes such as *public verifiability* and *forward secrecy of message confidentiality* while the others do not provide them. Some signcryption schemes are based on elliptic curve cryptography while the others use traditional approaches. The elliptic curve-based approaches can attain to a desired security level with significantly smaller keys that lead to efficient use of power, bandwidth, and storage that are basic limitations for resource-constrained environments.

In this paper, an elliptic curve-based signcryption scheme is introduced that provides the attributes of message confidentiality, authentication, integrity, unforgeability, non-repudiation, direct public verifiability, and forward secrecy of message confidentiality. The rest of this paper is organized as follows. Our proposed scheme is presented at Section 2 and its security attributes are analyzed in Section 3. Section 4 considers the computational costs, and Section 5 concludes the paper.

## 2. The Proposed Scheme

The proposed signcryption scheme consists of three phases: Initialization, Signcryption, and Unsigncryption that are described in following sections. Throughout this paper, *Alice* is sender, *Bob* is recipient, and *Mallory* is the malicious active attacker.

### 2.1. Initialization

Domain parameters of the proposed scheme consist of a suitably selected elliptic curve $E$ defined over a finite field $F_q$ with the Weierstrass equation of the form $y^2 = x^3 + ax + b$ and a *base point* $G \in E(F_q)$ in which $q$ is a large prime number. In order to make the elliptic curve non-singular, $a, b \in F_q$ should satisfy $4a^3 + 27b^2 \neq 0 \pmod{q}$ [8]. To guard against *small subgroup attacks*, the point $G$ should be of a prime order $n$ or equivalently, $nG = O$ where $O$ denotes the point of elliptic curve at infinity, and we should have $n > 4\sqrt{q}$ [9]. To protect against other known attacks on special classes of elliptic curves, $n$ should not divide $q^i - 1$ for all $1 \leq i \leq V$ ($V = 20$ suffices in practice), $n \neq q$ should be satisfied, and the curve should be non-supersingular [9]. To retain the intractability of ECDLP, $n$ should at least satisfy $n > 2^{160}$ for the common applications.

The private keys of *Alice* and *Bob* are randomly selected integers $w_A, w_B \in_R [1, n-1]$ and their public keys are calculated as $W_A = w_A G$ and $W_B = w_B G$ respectively. *Alice* and *Bob* are uniquely identified by their unique identifiers $ID_A$ and $ID_B$. They get digital certificates $Cert_A$ and $Cert_B$ for their public keys from the *Certificate Authority* (*CA*). If *CA* is not involved in the public key generation, it is necessary for *CA* to verify that each entity really possesses the corresponding private key of its claimed public key. This can be accomplished by a zero-knowledge technique. It should also be verified that the public keys belong to the main group.

---

[‡] Corresponding Author, ResearcherID: A-9528-2009

## 2.2. Signcryption

A schema of signcryption and unsigncryption stages of the proposed scheme is depicted in Figure 1. *Alice* generates the signcrypted text $(R,C,s)$ by following the below steps:

(1) Checks the validity of $Cert_B$ and uses it for verifying $W_B$. The process of certificate validation includes [10]:
  - Verifying the integrity and authenticity of the certificate by verifying the *CA*'s signature on the certificate.
  - Verifying that the certificate is not expired.
  - Verifying that the certificate is not revoked.
(2) Randomly selects an integer $r \in_R [1, n-1]$.
(3) Computes $R = rG$ where $R = (x_R, y_R)$ in which $x_R / y_R$ denotes the *x/y*-coordinate of the point *R*.
(4) Computes $K = (r + \tilde{x}_R w_A)W_B$ where $K = (x_K, y_K)$, and $\tilde{x}_R = 2^{\lceil f/2 \rceil} + (x_R \bmod 2^{\lceil f/2 \rceil})$ in which $f = \lfloor \log_2 n \rfloor + 1$ is the bit length of *n*, $\lfloor . \rfloor$ denotes the floor, and $\lceil . \rceil$ indicates the ceiling. If $K = O$ she goes back to step 2. Otherwise, she drives the session key of encryption as $k = H(x_K \| ID_A \| y_K \| ID_B)$ in which *H* is a one-way hash function that generates the required number of bits as the secret key of deployed symmetric encryption algorithm, and $\|$ denotes the concatenation.
(5) Computes the ciphertext as $C = E_k(M)$ in which $E_k(.)$ denotes a strong symmetric encryption algorithm (e.g. AES) that uses session key *k* for the encryption.
(6) Computes the digital signature as $s = tw_A - r \pmod{n}$ in which $t = H(C \| x_R \| ID_A \| y_R \| ID_B)$.
(7) Sends the signcrypted text $(R,C,s)$ to *Bob*.

## 2.3. Unsigncryption

*Bob* who received the signcrypted text $(R,C,s)$, follows the below steps to extract the plaintext and verify the signature:

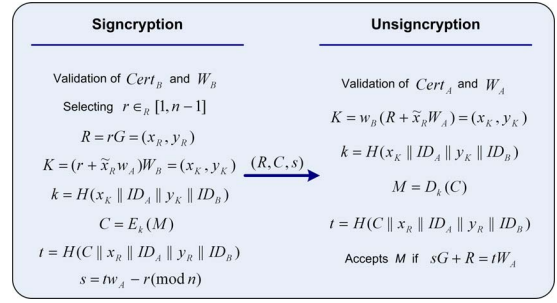

**Figure 1. The proposed scheme**

(1) Checks the validity of $Cert_A$ and uses it for verifying $W_A$.
(2) Computes $K = w_B(R + \tilde{x}_R W_A) = (x_K, y_K)$ and derives the session key as $k = H(x_K \| ID_A \| y_K \| ID_B)$.
(3) Decrypts the ciphertext as $M = D_k(C)$.
(4) Computes $t = H(C \| x_R \| ID_A \| y_R \| ID_B)$.
(5) Accepts *M* as the correct plaintext of *Alice* if and only if $sG + R = tW_A$. Otherwise, he rejects *M*.

## 3. Security Analysis

The correctness of the proposed scheme can be simply verified since *Alice* and *Bob* reach to the same point *K* on the elliptic curve:

$$K_A = (r + \tilde{x}_R w_A)W_B = (r + \tilde{x}_R w_A)w_B G = \\ = w_B(R + \tilde{x}_R W_A) = K_B = (x_K, y_K) \quad (1)$$

Defining $\tilde{x}_R$ as the least significant half in binary representation of $x_R$ is just a trade-off between security and efficiency. The proposed scheme deploys a strong key establishment. It tries to improve and match ideas behind MQV [11] and HMQV [12] protocols for its own case. The session key establishment part of the proposed scheme has itself the following security attributes:

a. *Known session key security:* Each message is signcrypted with a unique session key since random number *r* is used for session key establishment. The session key will also differ for different recipients since their public keys are involved in key derivation function.
b. *Resilience to Unknown-Key Share (UKS) attack:* The UKS attack [9] is thwarted in the proposed scheme since identifiers of both *Alice* and *Bob* are involved in the session key derivation function.
c. *Resilience to Key Compromise Impersonation (KCI) attack:* Under intractability of ECDLP, the KCI attack [9] is thwarted in the proposed scheme. An adversary that could obtain $w_A$ should find the corresponding *r* of *R* in order to deduce the corresponding session key that is generally in deposit of solving the ECDLP.

d. *Partial Forward secrecy:* Session key derivation function of the proposed scheme provides *partial forward secrecy* since even if $w_A$ is revealed, finding the corresponding random number $r$ is still necessary that is generally in deposit of solving the ECDLP.

Table 1 shows the provided attributes of the proposed scheme and those of other schemes. Hereunder, a brief proof is given for the claimed attributes.

1) *Confidentiality*: The proposed scheme is supposed to deploy a strong block cipher so the secrecy resides in the established session key. As it was described, the session key establishment provides several security attributes. Ultimately, an adversary has only two ways to defeat the confidentiality: having $w_B$, or deriving both $w_A$ and $r$ that are in deposit of solving the ECDLP.

2) *Authentication*: The proposed scheme is certificate-based, and certificates are verified by both sender and recipient. An implicit authentication is also involved in session key establishment so only the correct party that has the true private key can reach to correct key agreement and perform the unsigncryption. An authentication is also accomplished when one verifies the signcryption.

3) *Unforgeability*: *Mallory* cannot forge the valid $(C, R, s)$ with his malicious $(C', R, s')$. A valid forged signature $s'$ should satisfy $s' = s + (t' - t)w_A$ so knowledge of $w_A$ is necessary. Otherwise, he cannot truly forge the signature and the forged signature will be simply recognized when verifying the signcryption.

4) *Non-repudiation*: This can be deduced from unforgeability. It is computationally infeasible to forge the signature without having $w_A$.

5) *Integrity*: The hash value of ciphertext concatenated with some variable parameters is involved in signcryption so it changes with any alteration in the plaintext. The integrity is guaranteed by security attributes of hash function and unforgeability of the signature. *Mallory* should also have the valid session key to encrypt his modified message. Otherwise, the altered message will not be correctly decrypted by *Bob*. The integrity is also implicitly verified when verifying the signcryption.

6) *Forward secrecy of message confidentiality*: As a one-pass scheme and without any session-specific input from *Bob*, we cannot prospect the proposed scheme for the *Perfect Forward Secrecy* but it provides the partial forward secrecy under intractability of the ECDLP. It means that even if $w_A$ is revealed, the adversary should not be capable of decrypting the previously signcrypted texts. It is an obvious attribute of the proposed scheme since if one wants to defeat the confidentiality, it is necessary to have the corresponding random number $r$.

**Table 1. A comparison between provided attributes of different signcryption schemes**

| Signcryption Schemes | Direct Public Verifiability | Confidentiality | Integrity | Non-repudiation | Unforgeability | Forward Secrecy |
|---|---|---|---|---|---|---|
| Zheng [1] | No | Yes | Yes | Using another Protocol | Yes | No |
| Jung et al. [2] | No | Yes | Yes | Using another Protocol | Yes | Yes |
| Zheng and Imai [3] | No | Yes | Yes | Using another Protocol | Yes | No |
| Bao and Deng [4] | No | Yes | Yes | Directly | Yes | No |
| Gamage et al. [5] | Yes | Yes | Yes | Directly | Yes | No |
| Han et al. [6] | No | No | No | Directly | No | No |
| Hwang et al. [7] | No | No | No | Directly | No | No |
| The proposed Scheme | Yes | Yes | Yes | Directly | Yes | Yes |

**Table 2. Required number of field operations for different signcryption schemes**

| Signcryption schemes | Participant | Exp | Div | ECPM | ECPA | Mul | Add | Hash |
|---|---|---|---|---|---|---|---|---|
| Zheng [1] | Alice | 1 | 1 | - | - | - | 1 | 2 |
| | Bob | 2 | - | - | - | 2 | - | 2 |
| Jung et al. [2] | Alice | 2 | 1 | - | - | - | 1 | 2 |
| | Bob | 3 | - | - | - | 1 | - | 2 |
| Bao and Deng [4] | Alice | 2 | 1 | - | - | - | 1 | 3 |
| | Bob | 3 | - | - | - | 1 | - | 3 |
| Gamage et al. [5] | Alice | 2 | 1 | - | - | - | 1 | 2 |
| | Bob | 3 | - | - | - | 1 | - | 2 |
| Zheng and Imai [3] | Alice | - | 1 | 1 | - | 1 | 1 | 2 |
| | Bob | - | - | 2 | 1 | 2 | - | 2 |
| Han et al. [6] | Alice | - | 1 | 2 | - | 2 | 1 | 2 |
| | Bob | - | 1 | 3 | 1 | 2 | - | 2 |
| Hwang et al. [7] | Alice | - | - | 2 | - | 1 | 1 | 1 |
| | Bob | - | - | 3 | 1 | - | - | 1 |
| The proposed Scheme | Alice | - | - | 2 | - | 2 | 2 | 2 |
| | Bob | - | - | 4 | 2 | - | - | 2 |

Exp (modular Exponentiation), Div (modular Division/inverse), ECPM (Elliptic Curve Point Multiplication), ECPA (Elliptic Curve Point Addition), Mul (modular Multiplication), Add (modular Addition), Hash (one-way Hash function).

7) *Direct Public verifiability*: Anyone who observes the transmitted $(R, C, s)$ can compute $t = H(C \| x_R \| ID_A \| y_R \| ID_B)$ and directly verify the signcryption by checking $sG + R = tW_A$.

## 4. Computational Costs

Table 2 shows the computational costs of different schemes including the proposed scheme with respect to different kinds of required operations, in which the computational costs of verifications and symmetric encryption are neglected. Let $\zeta = \lfloor \log_2 n \rfloor + 1$ denotes the bit-length of modulus $n$. Using the conventional methods, the running time for modular addition and subtraction is of $O(\zeta)$ while it is of $O(\zeta^2)$ for modular multiplication and division, and is of $O(\zeta^3)$ for modular exponentiation and inverse calculation [13]. Total number of operations for calculating SHA-1 and MD5 hash functions are calculated as 1110 and 744 bit operations respectively [14]. Elliptic curve point multiplication is a time-consuming operation for which several methods are proposed in literature but selecting the efficient algorithm is complicated by many factors [8]. For the case of elliptic curve P-192 defined over $GF(p = 2^{192} - 2^{64} - 1)$, which provides the same level of security as that of RSA with 1024-bits modulus, total required number of operations for a point multiplication, for the case of an unknown point using the *Window NAF* method on Jacobian-Chudnovsky coordinate is reported in [8] as:

$$T_{ECPM}|_{\text{Jacobian-Chudnovsky}} = 1936 T_{Mul} + T_{Inv} \qquad (2)$$

For the same curve, the computational cost of an elliptic curve point addition in Jacobian projective coordinate [8] is:

$$T_{ECPA}|_{\text{Jacobian}} = 16 T_{Mul} + 7 T_{Add} \qquad (3)$$

The computational costs of the proposed scheme can be easily compared with those of other schemes presented in Table 2 by calculating total required number of operations, as it is accomplished in Figure 2. Figure 2 shows that the proposed scheme has a great computational advantage over exponentiation-based schemes while it provides the highest number of security attributes, as it is described in Table 1.

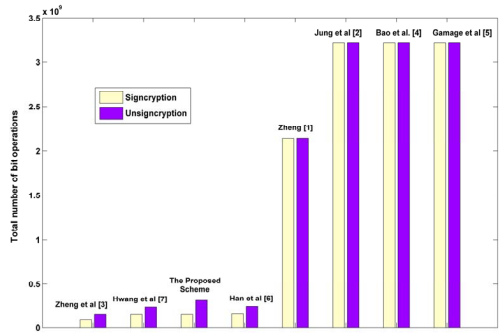

**Figure 2. Computational costs of different signcryption schemes using conventional methods**

## 5. Conclusions

A directly public verifiable signcryption scheme based on elliptic curves is introduced in this paper that provides the attributes of message confidentiality, authentication, integrity, unforgeability, and non-repudiation. It provides the forward secrecy of message confidentiality so even if the sender's private key is revealed, no one else can extract the plaintext of the previously signcrypted texts. It is directly public verifiable so anyone can verify the signature without any need for any secret information from the corresponding participants. It has great computational advantages over traditional schemes, and is so suitable for resource-constrained environments.

## References

[1] Y. Zheng, "Digital signcryption or how to achieve Cost (Signature & Encryption) << Cost (Signature) + Cost (Encryption)," Advances in Cryptology–CRYPTO'97, LNCS 1294, pp.165-179, Springer-Verlag, 1997.
[2] H.Y. Jung, K.S. Chang, D.H. Lee, and J.I. Lim, "Signcryption schemes with forward secrecy," Proceeding of Information Security Application-WISA 2001, pp.403-475, 2001.
[3] Y. Zheng, and H. Imai, "How to construct efficient signcryption schemes on elliptic curves," Information Processing Letters, Vol.68, pp.227-233, Elsevier, 1998.
[4] F. Bao, and R.H. Deng, "A signcryption scheme with signature directly verifiable by public key," Advances in Cryptology–PKC'98, LNCS 1431, pp.55-59, Springer-Verlag, 1998.
[5] C. Gamage, J. Leiwo, and Y. Zheng, "Encrypted message authentication by firewalls," International Workshop on Practice and Theory in Public Key Cryptography (PKC-99), LNCS 1560, pp.69-81, Springer-Verlag, March 1999.
[6] Y. Han, X. Yang, and Y. Hu, "Signcryption Based on Elliptic Curve and Its Multi-Party Schemes", 3rd ACM International Conference on Information Security (InfoSecu'04), pp.216-217, 2004.
[7] R.-J. Hwang, C.-H. Lai, and F.-F. Su, "An efficient signcryption scheme with forward secrecy based on elliptic curve," Journal of Applied Mathematics and Computation, Vol.167, No.2, pp.870-881, Elsevier, 2005.
[8] D. Hankerson, A. Menezes, and Scott Vanstone, "Guide to Elliptic Curve Cryptography," Springer-Verlag, New York, 2004.
[9] M. Toorani, and A.A. Beheshti Shirazi, "Cryptanalysis of an efficient signcryption scheme with forward secrecy based on elliptic curve," Proceedings of 2008 International Conference on Computer and Electrical Engineering (ICCEE'08), pp.428-432, Thailand, Dec. 2008.
[10] D.R. Stinson, "Cryptography-Theory and Practice," 3rd edition, Chapman & Hall/CRC, 2006.
[11] L. Law, A. Menezes, M. Qu, J. Solinas, and S. Vanstone, "An efficient Protocol for Authenticated Key Agreement", Journal of Designs, Codes and Cryptography, Vol.28, pp.119-134, 2003.
[12] H. Krawczyk, "HMQV: A high-performance secure Diffie-Hellman protocol (Extended Abstract)," Advances in Cryptology – CRYPTO'05, LNCS 3621, pp.546-566, Springer-Verlag, 2005.
[13] K.H. Rosen, "Elementary Number Theory and Its Applications," 2nd edition, Addison-Wesley, 1988.
[14] O. Elkeelany, M.M. Matalgah, K.P. Sheikh, M. Thaker, G. Chaudhry, D. Medhi, and J. Qaddour, "Performance Analysis of IPSec Protocol: Encryption and Authentication," Proceedings of the IEEE International Conference on Communications, Vol.2, pp.1164-1168, 2002.